# On the importance and challenges of modelling extraterrestrial photopigments via density-functional theory

**Dorothea Illner,[1] Manasvi Lingam,[2] and Roberto Peverati[1*]**

[1] Department of Chemistry and Chemical Engineering, Florida Institute of Technology, Melbourne, FL - 32901, USA
[2] Department of Aerospace, Physics, and Space Sciences, Florida Institute of Technology, Melbourne, FL - 32901, USA

[*]Corresponding author e-mail: rpeverati@fit.edu

## Abstract

The emergence of oxygenic photosynthesis was a major event in Earth's evolutionary history and was facilitated by chlorophylls (a major category of photopigments). The accurate modelling of photopigments is important to understand the characteristics of putative extraterrestrial life and its spectral signatures (detectable by future telescopes). In this paper, we perform a detailed assessment of various time-dependent density-functional theory (TD-DFT) methods for predicting the absorption spectra of chlorophyll *a*, with particular emphasis on modern low-cost approximations. We also investigate a potential extraterrestrial photopigment called phot0 and demonstrate that the electronegativity of the metal ion may exert a direct influence on the locations of the absorption peaks, with higher electronegativity inducing blue-shifting and vice-versa. Based on these calculations, we established that global-hybrid approximations with a moderate percentage of exact exchange – such as M06 and PW6B95 – are the most appropriate compromise between cost and accuracy for the computational characterization of photopigments of astrobiological interest. We conclude with a brief assessment of the implications and avenues for future research.

## 1. Introduction

It is well-established that photosynthesis is a key cornerstone of life on Earth, both in terrestrial and aquatic ecosystems [1]–[3]. For instance, approximately 80% of all biomass on Earth occurs as photosynthetic organisms, namely, embryophytes (land plants) [4]. Given that photosynthesis requires access to sunlight and (relatively) simple compounds for its operation (i.e., biosynthesis of organic compounds and energy transduction), it is not surprising that the earliest markers of photosynthesis seem to manifest not long after (i.e., a few 100 Myr) the earliest traces of life on Earth [1],[5]–[8].

Among the many variants of photosynthesis documented, oxygenic photosynthesis – which entails the production of molecular oxygen – dominates in terms of biomass [4]. In oxygenic photosynthesis, the electron donor is water, which has been abundant on Earth. In contrast, anoxygenic photosynthesis involves electron donors such as ferrous iron and hydrogen sulfide [9]. Limited access to these electron donors may have constrained the extent of photosynthesis in the Archean eon [10], whereas this potential bottleneck ought not to be an issue for oxygenic photosynthesis, thus possibly serving to partly explain its ubiquity.

The biological advantages of oxygenic photosynthesis, especially in connection with its production of oxygen and the subsequent oxygenation of the atmosphere, are well documented. The rise in oxygen levels permitted the formation of an ozone layer, triggered the formation of new minerals, created new ecological niches, and might have partially aided in the evolution of complex, motile, and macroscopic multicellular life [7],[8],[11],[12].

In view of all these advantages, it is apparent that gaining a theoretical understanding of oxygenic photosynthesis – hereafter labelled as ‘photosynthesis’ for simplicity – is beneficial. Photopigments are vital components of photosynthesis, as they enable the transduction of light energy (into chemical energy); among the array of photopigments, the chlorophylls (Chls) are integral for energy harvesting [1]. The importance of Chls extends beyond the biological sciences because they are being extensively investigated in artificial photosynthesis, a rapidly growing discipline in renewable energy [13]–[16].

However, in the context of this paper, we shall highlight the importance of photosynthesis in another domain, namely, astrobiology – specifically in the swiftly expanding discipline of exoplanets [17]–[19]. As summarised in the next section, there is growing evidence that an improved understanding and modelling of photosynthesis may be valuable for settling the fundamental question: *Are we alone?*

The outline of the paper is as follows. In section 2, we rationalize the spectra of chlorophyll and other photopigments, and we provide a brief primer to the research undertaken in exoplanetary science on photosynthesis. In section 3, we present our computational results on the performance of several modern density functional theory (DFT) methods for calculations of absorption spectra of chlorophyll *a* and other photopigments. Finally, in section 4, we provide recommendations concerning appropriate computational protocols and a perspective on the importance and challenges of using DFT calculations for exoplanetary research.

# 2. Basics of photosynthesis in terrestrial and exoplanetary environments

## 2.1 Rationalization of the absorption spectra of chlorophylls and other photopigments

The spectroscopic properties of Chls and other similar photopigments can be rationalized, to leading degree, by using the model introduced by Gouterman to explain the absorption spectra of porphyrins [20]. This model involves excitations between four frontier orbitals: HOMO–1, HOMO, LUMO, and LUMO+1. In an idealized $D_{4h}$ symmetry, the two occupied orbitals transform as the accidentally degenerate $a_{2u}$ and $a_{1u}$ pair, while the two unoccupied ones transform as the set of degenerate $e_g$ orbitals. Transitions between these orbitals result in two excited states, both of $^1E_u$ character, split in energy by orbital mixing. The two ensuing absorption bands are: i) a relatively high-intensity B-band (sometimes also called Soret band) at lower wavelengths, corresponding to the pair of transitions with greater excitation energy and oscillator strength, and ii) a relatively low-intensity Q-band at higher wavelengths, corresponding to the pair of transitions with lower energy and oscillator strength.

Since chlorophylls break the idealized $D_{4h}$ symmetry, all degeneracies between orbitals are lifted, but a qualitative resemblance to the idealized $D_{4h}$ case is maintained and is reflected in their spectra. Namely, the Q- and B-bands remain well separated, each splitting into two distinctly identifiable transitions. The transitions can be labelled according to the polarization direction within the macrocycle plane, which, in order of increasing energy, are: $Q_y$, $Q_x$, and $B_x$, $B_y$. In the idealised case, the *y*-labelled excitations should primarily correspond to the HOMO → LUMO and HOMO–1 → LUMO+1 transitions, while the *x*-labelled ones should correspond to the HOMO–1 → LUMO and HOMO → LUMO+1 transitions, see for example Figure 3 of Ref. [21].

The energies of the frontier orbitals of photopigments can be shifted by functionalizing the central ring system (resembling chlorin) with substituents or changing the central metal. The former is the preferred tool of evolution in a chemically rich environment, such as on Earth. This is evidenced by many naturally occurring Chls, which differ from each other by presenting substituents at eight locations on the chlorophyll ring [22]. At a more primordial stage of evolution, however, it is conceivable that replacing the central metal might be one of the most straightforward methods for modulating the absorption properties of primitive photopigments. For example, calcium and zinc possess electronic properties similar to magnesium and are relatively less abundant, but still quite common, in the Universe [23]. The main effect of a metal on the orbital energies is through conjugation of its $p_\pi$ orbital with the π system of the photopigment. Symmetry consideration restricts the interaction of the metal with one orbital, which is usually the $a_{2u}$ orbital in an idealized $D_{4h}$ molecule (or the $a_2$ orbital in the $C_{2v}$ symmetry of Chl *a*) [20]. In this simplified view of the interaction, the electronegativity of the metal plays a crucial role in raising or lowering the orbital energies [20],[24],[25]: A more electronegative metal will raise the energy of the orbital, while a more electropositive one will lower it.

### 2.2 Extraterrestrial photosynthesis: A primer

Section 1 has highlighted the biological importance of (oxygenic) photosynthesis on Earth. Since we know of merely one world with life (i.e., our planet), a common strategy in astrobiology is to use the Earth as a starting point and extrapolate accordingly.

By this logic, one could theorise that photosynthesis may be operational on other worlds. If this were to be correct, the immediate question that springs to mind is: *how can photosynthesis be detected, in principle, by the next generation of telescopes?* Two major biological signatures (biosignatures) of photosynthesis are considered 'canonical' in astrobiology:

1. Molecular oxygen ($O_2$): As noted previously, $O_2$ is a metabolic product of photosynthesis. In principle, it can be detected spectroscopically by searching for absorption features such as the $O_2$-A band (~760 nm) and the $O_2$-B band (~690 nm) [18]. However, aside from possible issues with its detectability [26], a major drawback of $O_2$ is that it is readily susceptible to false positives, i.e., it can be produced by abiotic processes [27],[28]. Hence, due care is needed when interpreting spectral signatures of $O_2$ in exoplanetary atmospheres.

2. Vegetation Red Edge (VRE): An interesting feature of most Chls is that their absorbance declines conspicuously beyond the $Q_y$ absorption band at ~650-700 nm (see Figure 4.8 of Ref. [1]), consequently translating to a sharp increase in reflectance at roughly 700 nm. This spectral feature, known as the VRE [29],[30], is considered a robust biosignature because it has no known false positives, and photopigments (to which Chls belong) are canonically perceived as reliable signatures of extraterrestrial life, i.e., their likelihood of abiotic synthesis is low [31]. The VRE is believed to have been discernible for as much as ~50% of Earth's geological history [32], and numerical simulations of reflected light spectra indicate that it may be detectable on exoplanets by forthcoming telescopes [33],[34].

However, an attendant subtlety concerning point #2 speaks to a broader debate in astrobiology about the use of Earth-based life as a proxy for extraterrestrial life. It is often implicitly assumed in publications pertaining to point #2 that putative photopigments on other worlds would be identical to Chls. It is important to recognise that this assumption is not automatically guaranteed, to wit, other worlds might host photopigments that diverge from Chls in some notable respects.

The majority of publications centred on exploring extraterrestrial (oxygenic) photosynthesis have tended to adopt Chls as the basis of their experimental or theoretical analyses [35]–[46]. On the other hand, there is growing dissemination of the notion that extraterrestrial photosynthesis, in general, and photopigments, in particular, may differ substantially from Earth [47]–[57]. To offer a specific example, some publications have theorised that the number of functional units of photosynthesis (photosystems) coupled together could be higher on planets orbiting cooler stars than the Sun, thereby allowing for longer wavelengths to be efficiently harvested [35],[38],[47],[48],[50]–[52].

In addition to such modifications, the absorption characteristics of the photopigments may themselves be modulated by planetary and stellar factors (e.g., star temperature). It was proposed in Ref. [48] that the absorption maxima of photopigments analogous to the Q-bands in Chls might be fairly proximal to the peak spectral photon flux of the host star; the corresponding wavelength ($\lambda_{opt}$) associated with the latter is given by $\lambda_{opt} \cong 635\ \mathrm{nm}\ (T_\star/5780\ \mathrm{K})$, where $T_\star$ is the blackbody temperature of the star [8]. Other papers have, likewise, sought to use criteria such as energy input maximisation or noise minimisation to predict absorption maxima of putative photopigments on exoplanets [38],[54],[56],[57].

A related strategy is to postulate that the molecular structures of the photopigments could diverge from Chls [52],[53],[55],[56]. For instance, Chls consist of a magnesium ion at the centre of the porphyrin ring system [1]. Therefore, replacing the magnesium with calcium (from the same group) or one of the period 4 transition metals that can exhibit the same closed shell (i.e., singlet) structure yields potential photopigments, as hypothesised in Refs. [52],[55],[56], which may accordingly affect the location of the absorption peaks compared to Chls.

Thus, on the basis of our preceding discussion, a number of reasons exist for a computational approach to the study of the absorption spectra of alternative photopigments:

1. This modelling could aid us in estimating the appropriate wavelength range needed for future observations, insofar as searching for spectral signatures of modified VREs are concerned.
2. This modelling may permit us to constrain what types of molecular structures (of photopigments) are compatible with absorption spectra predicted by the aforementioned noise and/or energy extremization theoretical models [38],[54],[56],[57], which are not capable of identifying such potential molecular structures.
3. Recent developments in astrophysics have shown that exoplanets have very diverse chemical compositions [58],[59], which is expected due to the heterogeneity of the pathways producing the elements [60],[61]. Hence, it is conceivable that certain elements (say, zinc or calcium) are more/less plentiful on some exoplanets and might thus be more/less easily assimilated into photopigments, consequently shaping the latter's absorption spectra.

Based on these considerations, we will consider two distinct photopigments – chlorophyll a (Chl *a*) and phot0, which represents a simplified photopigment proposed by Ref. [55] as a plausible precursor to Chls – and replace the magnesium ion at their centres with other metals in the subsequent Sections.

### 2.3 Experimental and computational approaches to chlorophylls spectra

Experimental spectra of Chls in gas-phase are challenging to obtain because neutral Chls readily decompose. Therefore, discussions of Chls absorption properties have historically been based on spectra obtained in solution (refer to Ref. [62] for relevant references). Early computational modelings of Chls have mainly focused on accurate calculations of the low-lying excited states of Chls [63]–[69]. Those calculations usually used data from the spectra in solution as general guidance rather than as a benchmark since some of the issues caused by the interaction with solvent molecules [70],[71] were generally neglected.

Recently, accurate gas-phase spectra of charged Chls became available with action spectroscopy by tagging the molecules with cations [72]–[75]. These experiments present their own set of issues, such as the effects of the tagging cation on the Soret region of the spectrum [75]. Combining experimental data from tetramethylammonium-tagged chlorophyll a from Ref. [74] with high-level calculations using the domain-based local pair natural orbital implementation of the similarity transformed equation of motion coupled cluster theory with single and double excitations (DLPNO-STEOM-CCSD), Sirohiwal et al. recently reported energy shifts to adjust the experimental absorption maxima to vertical excitation energies (VEEs) [21]. With these shifts, they report 'quasi-experimental VEEs' for Chl *a*, which can be used to directly compare calculated VEEs with the gas-phase spectra [76]. In the following sections, we use these values as the most accurate benchmark for Chl *a*. Additionally, we complement these results for Chl *a* with our own DLPNO-STEOM-CCSD calculations to generate data for molecules where experimental spectra are unavailable.

Several high-level methodologies have been used to calculate the low-lying energy spectrum of Chl *a,* and results are available in the literature. The highly-accurate DLPNO-STEOM-CCSD results of Sirohiwal *et al.* align with previously reported SAC-CI calculations for both Q- and B-bands [63],[69]. In addition, their results for the Q-band are also in line with other coupled-cluster approximations, such as CC2 and ADC(2) [77]. However, time-dependent density-functional theory (TD-DFT) is usually the method of choice for the calculation of absorption spectra of medium to large photopigments because of the favourable compromise between accuracy and computational cost. Recent TD-DFT comparisons with high-level data show that none of the exchange–correlation functionals that have been tested to date "*can provide a one-stop-solution, in the framework of TD-DFT, for all low-energy absorption features of Chl a*" [21], except for the B2PLYP double-hybrid functional. The main drawback of TD-DFT calculations is the well-known limitations of approximated functionals to accurately describe charge-transfer (CT) states, which results in both i) a blue-shift of Gouterman states energies and ii) the overstabilization of non-Gouterman states with significant CT character. The issue of TD-DFT with CT states is documented in the literature well beyond chlorophylls [78]–[81]. This limitation is usually attributed to self-interaction errors, which can be mitigated by including a portion of exact exchange with global- or range–separated-hybrid functionals. Application of such functionals to chlorophylls, however, has not been as successful as in other cases. In the following section, we expand these comparisons to a larger number of previously untested exchange–correlation functionals, with a particular emphasis on range-separated ones. In particular, we aimed to test the performance of many modern functionals that could provide results comparable to double-hybrid functionals at a lower computational cost.

# 3. Results

## 3.1 Benchmarking modern time-dependent density-functional theory calculations of absorption spectra of chlorophyll *a*

In order to expand the benchmark of modern exchange–correlation approximations on the absorption spectra of Chl *a*, we performed TD-DFT calculations using the def2-TZVP basis set in conjunction with 25 previously untested functionals. All calculations have been performed with the Q-Chem 6.1 program [82], which includes a comprehensive list of exchange–correlation approximations. Among a plethora of more than 250 functionals available in the program, we selected all of the range-separated hybrids, as well as some functionals that have been singled out in recent literature as methods that perform well for a broad range of properties [83]–[89]. All results reported in the main text of this work use the same gas-phase molecular geometry calculated at the CAM-B3LYP-D3(BJ)/def2-TZVP level of theory and reported by Sirohiwal et al. in Ref. [21]. This geometry replaces the phytyl chain at position 17 of Chl *a* with methyl and is the smallest geometry that can significantly represent Chl *a*, containing only 73 atoms. VEEs of the Q- and B-bands of Chl *a* with all considered approximations are reported in Table 1.

**Table 1:** Vertical Excitation Energies (VEEs, in eV) associated with the Q- and B-bands of Chlorophyll *a* calculated using DLPNO-STEOM-CCSD and TD-DFT with various exchange–correlation functionals. Results are compared with quasi-experimental VEEs reported in Ref. [21]. The state number based on the rank of the root is reported for each state in parenthesis.

| **Method:** | **Type[a]** | **Reference** | $\mathbf{Q}_y$ | $\mathbf{Q}_x$ | $\mathbf{B}_x$ | $\mathbf{B}_y$ |
|---|---|---|---|---|---|---|
| **Quasi-experimental** | | | 1.99 (1) | 2.30 (2) | 3.12 (3) | 3.38 (4) |
| **DLPNO-STEOM-CCSD** | WFT | | 1.75 (1) | 2.24 (2) | 3.17 (3) | 3.40 (4) |
| BLYP | L | [90],[91] | 2.05 (1) | 2.13 (2) | 2.82 (6) | 2.95 (7) |
| PBE | L | [92] | 2.07 (1) | 2.14 (2) | 2.83 (6) | 2.96 (7) |
| BP86 | L | [90],[93] | 2.06 (1) | 2.14 (2) | 2.84 (6) | 2.96 (7) |
| M06-L | L | [94] | 2.13 (1) | 2.24 (2) | 3.01 (6) | 3.15 (7) |
| MN15-L | L | [95] | 2.23 (1) | 2.38 (2) | 3.17 (6) | 3.34 (7) |
| TPSS | L | [96] | 2.09 (1) | 2.18 (2) | 2.90 (6) | 3.03 (7) |
| SCAN | L | [97] | 2.13 (1) | 2.24 (2) | 3.02 (6) | 3.15 (7) |
| $r^2$SCAN | L | [98] | 2.16 (1) | 2.26 (2) | 3.01 (6) | 3.15 (7) |
| B97M-V | L | [99] | 2.15 (1) | 2.28 (2) | 3.05 (6) | 3.20 (7) |
| B3LYP | GH (20%) | [90],[91],[100] | 2.16 (1) | 2.35 (2) | 3.17 (4) | 3.35 (7) |
| PBE0 | GH (25%) | [101] | 2.19 (1) | 2.40 (2) | 3.26 (4) | 3.45 (7) |
| TPSSh | GH (10%) | [102] | 2.16 (1) | 2.29 (2) | 3.09 (6) | 3.24 (7) |
| $r^2$SCANh | GH (25%) | [103] | 2.20 (1) | 2.35 (2) | 3.17 (6) | 3.33 (7) |
| $r^2$SCAN0 | GH (25%) | [103] | 2.24 (1) | 2.47 (2) | 3.36 (4) | 3.56 (6) |
| PW6B95 | GH (28%) | [104] | 2.17 (1) | 2.40 (2) | 3.22 (3) | 3.27 (4) |
| M06 | GH (27%) | [105] | 2.11 (1) | 2.35 (2) | 3.23 (4) | 3.42 (7) |
| M06-2X | GH (54%) | [105] | 2.19 (1) | 2.56 (2) | 3.44 (3) | 3.69 (5) |
| M08-HX | GH (52%) | [106] | 2.21 (1) | 2.56 (2) | 3.44 (3) | 3.68 (5) |
| M08-SO | GH (57%) | [106] | 2.15 (1) | 2.54 (2) | 3.41 (3) | 3.68 (5) |
| MN15 | GH (44%) | [85] | 2.15 (1) | 2.47 (2) | 3.33 (3) | 3.56 (6) |
| LC-ωPBE | RSH (0–100%) | [107] | 2.07 (1) | 2.59 (2) | 3.47 (3) | 3.78 (5) |
| LC-ωPBEh | RSH (20–100%) | [108] | 2.12 (1) | 2.54 (2) | 3.44 (3) | 3.73 (5) |
| CAM-B3LYP | RSH (19–100%) | [109] | 2.14 (1) | 2.53 (2) | 3.43 (3) | 3.71 (5) |
| ωB97M-V | RSH (15–100%) | [110] | 2.06 (1) | 2.64 (2) | 3.49 (3) | 3.85 (5) |
| ωB97X-V | RSH (17–100%) | [111] | 2.07 (1) | 2.68 (2) | 3.54 (3) | 3.91 (5) |
| M11 | RSH (43–100%) | [112] | 2.11 (1) | 2.67 (2) | 3.52 (3) | 3.86 (6) |
| revM11 | RSH (23–100%) | [113] | 2.02 (1) | 2.75 (2) | 3.52 (3) | 3.89 (5) |
| ωB97X-D | RSH (22–100%) | [114] | 2.12 (1) | 2.55 (2) | 3.45 (3) | 3.75 (5) |
| ωB97X-D3 | RSH (20–100%) | [115] | 2.09 (1) | 2.61 (2) | 3.50 (3) | 3.83 (5) |
| ωM05-D | RSH (37–100%) | [116] | 2.11 (1) | 2.59 (2) | 3.47 (3) | 3.78 (5) |
| ωM06-D3 | RSH (27–100%) | [115] | 2.08 (1) | 2.69 (2) | 3.55 (3) | 3.88 (4) |
| B2PLYP | DH | [117] | 2.12 (1) | 2.23 (2) | 3.17 (3) | 3.27 (4) |
| ωB2PLYP | RS-DH | [118] | 2.04 (1) | 2.49 (2) | 3.45 (3) | 3.78 (4) |

[a]The type of functionals are reported in this column, with the following abbreviations: WFT=Wave Function Theory; L=Local; GH=Global-Hybrid; RSH=Range-Separated–Hybrid; DH=Double-Hybrid; RS-DH=Range-Separated Double-Hybrid. For GH and RSH functionals the percentage of exact exchange is also reported in parenthesis.

In the context of a computational investigation of photopigments in exoplanetary science, it is essential to accurately describe the absorption spectra, with a particular focus on the Q-band. From an astrobiological perspective, the increase in reflectance at wavelengths greater than the Q-band is directly linked to the VRE, as noted in Section 2.2. For such reason, we designed three statistical indicators that highlight the performance of the TD-DFT calculations along the following aspects: 1) the performance for the Q-band (Q-MUE), 2) the performance for the B-band (B-MUE), and 3) the presence of so-called 'ghost states' between them, which complicates the interpretation and assignment of the peaks (#GS). The Q-MUE and B-MUE indicators are mean unsigned errors (MUEs) calculated with respect to the quasi-experimental VEEs of Sirohiwal et al. as the reference results for the Q- and B-band, respectively (each band is composed of two Gouterman-type transitions). The #GS indicator is calculated by taking the average number of non-Gouterman states with energies between those of the Gouterman ones. These states are 'ghost states' with a substantial charge-transfer character that get over-stabilized by an incorrect asymptotic form of the exchange–correlation potential. The presence of these states is generally not a significant issue when spectra are computed since they usually have low oscillator strength. However, their presence requires calculations and interpretations of a higher number of states, which might complicate the analysis and increase the computational cost. Finally, we also combined the three statistical indicators into an overall weighted mean unsigned error (wMUE) using the following formula:

$$\mathrm{wMUE} = w_Q\ \mathrm{Q\text{-}MUE} + w_B\ \mathrm{Q\text{-}MUE} + w_G\ \#\mathrm{GS}, \quad (1)$$

with $w_Q = 1$, $w_B = 1/2$, and $w_G = 1/10$. While there is an intrinsic bias in the selection of the weights, we chose their values to provide a qualitative ranking in the context of the prediction of absorption spectra in exoplanetary conditions. These statistical indicators are reported in Figure 1.

| Method | Type | Q-MUE | B-MUE | #GS | wMUE |
|---|---|---|---|---|---|
| **B2PLYP** | DH | 0.10 | 0.08 | 0 | 0.140 |
| **DLPNO-STEOM-CCSD** | WFT | 0.15 | 0.03 | 0 | 0.168 |
| **PW6B95** | GH (28%) | 0.14 | 0.11 | 0 | 0.194 |
| **ωB2PLYP** | RS-DH | 0.12 | 0.37 | 0 | 0.303 |
| **M06** | GH (27%) | 0.09 | 0.08 | 2 | 0.323 |
| **B3LYP** | GH (20%) | 0.11 | 0.04 | 2 | 0.328 |
| **MN15** | GH (44%) | 0.16 | 0.20 | 1 | 0.362 |
| **M08-SO** | GH (57%) | 0.20 | 0.29 | 0.5 | 0.395 |
| **CAM-B3LYP** | RSH (19–100%) | 0.19 | 0.32 | 0.5 | 0.401 |
| **PBE0** | GH (25%) | 0.15 | 0.10 | 2 | 0.401 |
| **LC-ωPBEh** | RSH (20–100%) | 0.18 | 0.33 | 0.5 | 0.402 |
| **ωB97X-D** | RSH (17–100%) | 0.19 | 0.35 | 0.5 | 0.411 |
| **LC-ωPBE** | RSH (0–100%) | 0.18 | 0.37 | 0.5 | 0.421 |
| **TPSSh** | GH (10%) | 0.09 | 0.08 | 3 | 0.430 |
| **M06-2X** | GH (54%) | 0.23 | 0.32 | 0.5 | 0.436 |
| **M08-HX** | GH (52%) | 0.24 | 0.31 | 0.5 | 0.444 |
| **ωM05-D** | RSH (37–100%) | 0.21 | 0.38 | 0.5 | 0.446 |
| **B97M-V** | L | 0.09 | 0.12 | 3 | 0.453 |
| **r²SCANh** | GH (10%) | 0.13 | 0.05 | 3 | 0.457 |
| **r²SCAN0** | GH (25%) | 0.21 | 0.21 | 1.5 | 0.462 |
| **ωB97X-D3** | RSH (20–100%) | 0.21 | 0.42 | 0.5 | 0.464 |
| **ωB97M-V** | RSH (15–100%) | 0.21 | 0.42 | 0.5 | 0.465 |
| **ωM06-D3** | RSH (27–100%) | 0.24 | 0.46 | 0 | 0.475 |
| **SCAN** | L | 0.10 | 0.17 | 3 | 0.480 |
| **MN15-L** | L | 0.16 | 0.05 | 3 | 0.483 |
| **M06-L** | L | 0.10 | 0.17 | 3 | 0.486 |
| **r²SCAN** | L | 0.10 | 0.17 | 3 | 0.488 |
| **revM11** | RSH (23–100%) | 0.24 | 0.45 | 0.5 | 0.514 |
| **ωB97X-V** | RSH (17–100%) | 0.23 | 0.48 | 0.5 | 0.518 |
| **TPSS** | L | 0.11 | 0.29 | 3 | 0.554 |
| **M11** | RSH (43–100%) | 0.24 | 0.44 | 1 | 0.563 |
| **BP86** | L | 0.12 | 0.35 | 3 | 0.591 |
| **PBE** | L | 0.12 | 0.35 | 3 | 0.593 |
| **BLYP** | L | 0.12 | 0.36 | 3 | 0.597 |

**Fig. 1:** Summary of the mean unsigned errors of the exchange–correlation functionals for the calculation of the absorption spectra of chlorophyll *a* with respect to the reference quasi-experimental VEEs. The statistical indicators represent the following: Q-MUE: mean unsigned error for the 2 transitions of the Q-band; B-MUE: mean unsigned error for the 2 transitions of the B-band; #GS: the average number of over-stabilized CT 'ghost states' between the main signals; wMUE: the weighted MUE results calculated using Eq. 1. Each cell is shaded according to a color scale where green indicates the best performance, and red indicates the worst performance. Labels for the type of functionals are the same as in Table 1. Methods are ranked in order of increasing wMUE.

Our results show that three methods perform substantially better than all others: the B2PLYP double-hybrid functional, the DLPNO-STEOM-CCSD wave function method, and the PW6B95 global-hybrid functional. While the first two methods were already highlighted in Ref. [21], our results show that the PW6B95 functional provides a comparable level of performance. Interestingly, the highly-accurate DLPNO-STEOM-CCSD method performs slightly worse than the B2PLYP double hybrid, primarily because of a $Q_y$ transition red-shifted by 0.24 eV. For all other transitions, the DLPNO-STEOM-CCSD outperforms every other functional, as expected. Focusing on the Q-band exclusively, we notice that most of the functionals have Q-MUEs below 0.15 eV, showing that TD-DFT calculations are generally reliable for the low-lying energy states. Among these reliable functionals, we notice mostly local and global-hybrids with low to moderate percentages of exact exchange. Because of their higher self-interaction errors, however, these functionals have a large number of overstabilized CT ghost states. The performance for the B-band is also compromised, especially for local functionals, which in general are unreliable for these states. Diminishing the self-interaction errors by using a higher percentage of exact exchange or by range-separation proves to be a double-edged sword. In fact, only global-hybrids with a low to moderate percentage of exact exchange provide a balanced description of all aspects of the absorption spectra. Among these functionals, the best methods across the board are the already mentioned PW6B95, as well as B3LYP, M06, TPSSh, and $r^2$SCANh. Notice that the last four methods all have a percentage of exact exchange close to 25% and perform on par or better than B2PLYP, as long as the issue of ghost states is ignored. Among global-hybrid functionals with high percentage of exact exchange, the MN15 and M08-SO approximations stand out, achieving wMUE comparable to functionals with smaller percentages. Considering the other classes of functionals, CAM-B3LYP is the best range-separated hybrid method, while B97M-V is the best local one. Both these classes, however, suffer from severe issues for these types of calculations and are therefore not recommended. Functionals in the range-separated hybrid class tend to overestimate the energies of the relevant transition states, while functionals in the local class overstabilize a large number of CT ghost states. It is also interesting to notice that the addition of the kinetic energy density as an additional functional ingredient seems to be irrelevant for this case. No significant difference is evident between the results of the so-called 'rung-2' GGA approximation and those of 'rung-3' meta-GGA ones.

Finally, we also investigated two other features that can affect the accuracy and cost of TD-DFT calculations: the effects of changes in the molecular geometry and the use of the Tamm-Dancoff approximation [78]. Since some dependence of the results on the molecular geometry has been reported by Sirohiwal *et al.* [21], we also explored the possibility of using the full Chl *a* molecular geometry containing 135 atoms. To do so, we extracted the geometry from the crystal structure of photosystem II, and we optimized it in gas-phase with the GFN2-xtb method as implemented in the *xtb* program [119]. Analysing the full geometry results, reported in Table S1 in the supporting information, we notice that most general trends and overall results do not change. The first difference we noticed is that results depend somewhat on the vinyl group's rotation, as Sirohiwal et al. reported. Range-separated hybrid functionals seem to be particularly sensitive to this issue, with the Q-band that is blue-shifted by an average of 0.1 eV and the B-band that is blue-shifted by an average of 0.2 eV when the CAM-B3LYP geometry results are compared to the full geometry ones. Another significant difference is that the local functionals are heavily affected by changes in geometry. In this case, it is not an issue of shifts of the peaks but rather an issue with the ghost states. The drastically incorrect asymptotic behavior of the exchange–correlation potential for these functionals overstabilize several CT states where the charge is delocalized between the ring and the phytyl chain. For example, popular functionals such as PBE, BLYP, and TPSS predict three CT states between the $Q_x$ and $B_x$ transition, with the $B_y$ transition right above them, when the CAM-B3LYP geometry is used, but they predict nine CT states between $Q_x$ and $B_x$, and three CT states between $B_x$ and $B_y$ with the full geometry. These states were never observed in the past because previous studies have always been conducted on truncated phytyl chains. However, this large number of ghost states complicates the interpretation of the results drastically and increases the computational cost because at least 15 states need to be computed for each calculation. To avoid these issues, we generally discourage the use of local functionals for the calculation of absorption spectra of Chlorophylls.

Global hybrid functionals with a moderate percentage of exact exchange are far less affected by changes in geometry and remain the best compromise for these types of calculations. Finally, the Tamm–Dancoff approximation [78] (i.e., neglecting the **B** matrix in the TD-DFT eigenvalue problem) is a standard algorithm to speed up TD-DFT calculations. We investigated the use of the Tamm–Dancoff approximation and reported the main results in Table S2 in the supporting information. Unfortunately, results with the Tamm–Dancoff approximation are different from the full TD-DFT results, especially for hybrid functionals. As such, we suggest avoiding the Tamm-Dancoff approximation for the description of the absorption spectra of Chl *a*.

### 3.2 Searching for extraterrestrial photopigments: Red-shifted and blue-shifted phot0 analogues

In this section, we consider modifications to the phot0 molecule presented by de la Concepcion *et al.* in Ref. [55]. We picked this system because it has been designed as a possible precursor to extraterrestrial photopigments. In Ref. [55], both $Mg^{2+}$ and $Zn^{2+}$ were considered as metallic centres. In this work, we expanded the number of metallic centres by adding $Ca^{2+}$ (see the structures in Figure 2).

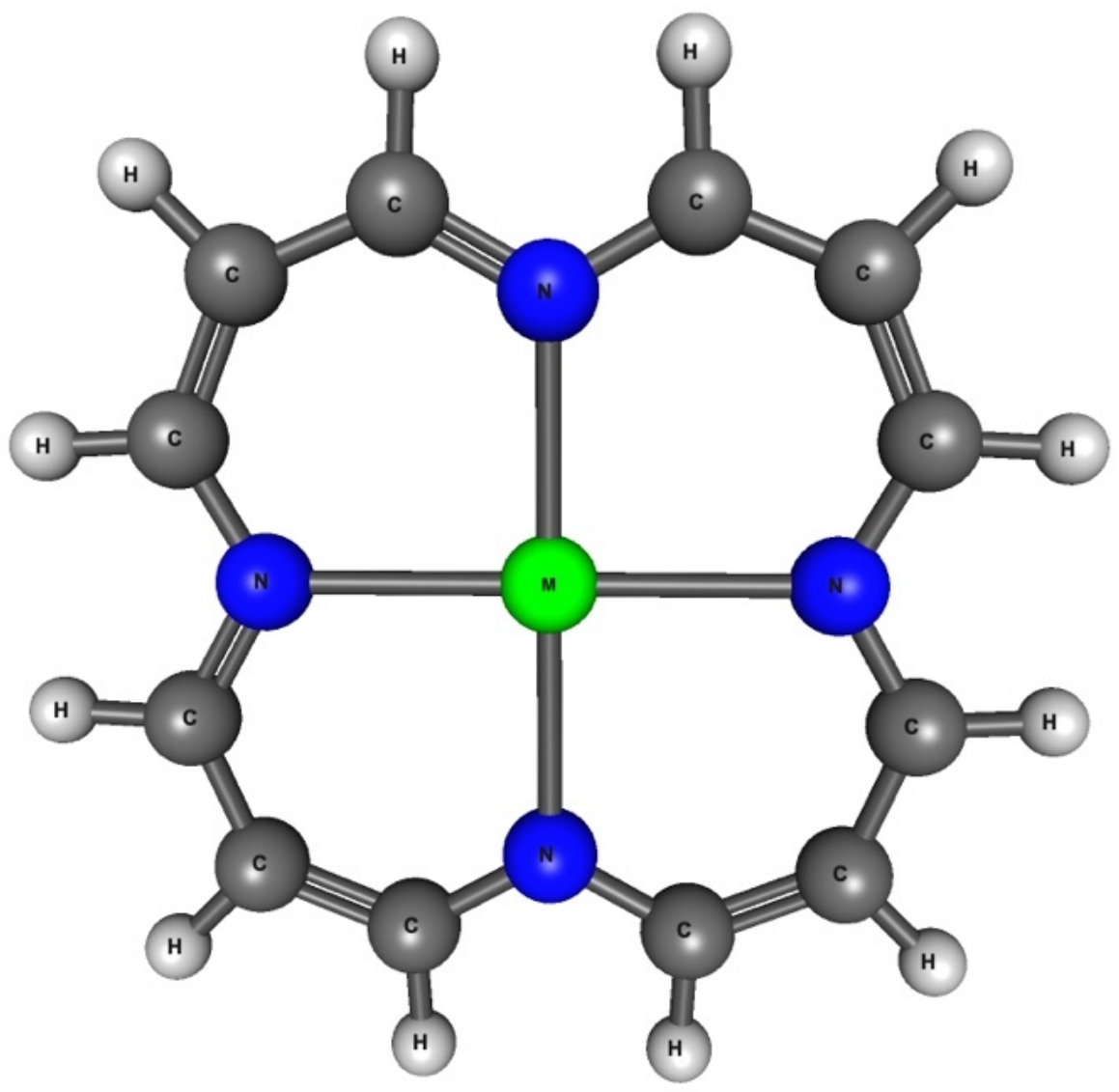

**Fig. 2:** General structure of the phot0 molecule. M is either $Ca^{2+}$ (phot0-Ca), or $Mg^{2+}$ (phot0-Mg), or $Zn^{2+}$ (phot0-Zn).

The choice of $Ca^{2+}$ is justified to study the effect of the metal on the absorption properties of the photopigment: Starting from Mg ($\chi_{Mg} = 1.3$) as the baseline in analogy with Chl *a*, we wanted to compare absorption spectra of phot0 with both a more electropositive metal ($\chi_{Ca} = 1.0$), and a more electronegative one ($\chi_{Zn} = 1.6$). We performed benchmark spectra calculations using the DLPNO-STEOM-CCSD method for all molecules on gas-phase geometries optimized at the CAM-B3LYP-D3(BJ)/def2-TZVP level of theory, in analogy with the previous section (provided in the supporting information). These calculations were performed using the ORCA 5.0.4 program [120], with the same setting used in Ref. [21] for Chl *a* (i.e., def2-TZVP basis set with tight convergence thresholds). All molecules have a closed-shell singlet ground state. Results are reported in Table 2, while simulated spectra using Lorentzian broadening are reported in Figure 3.

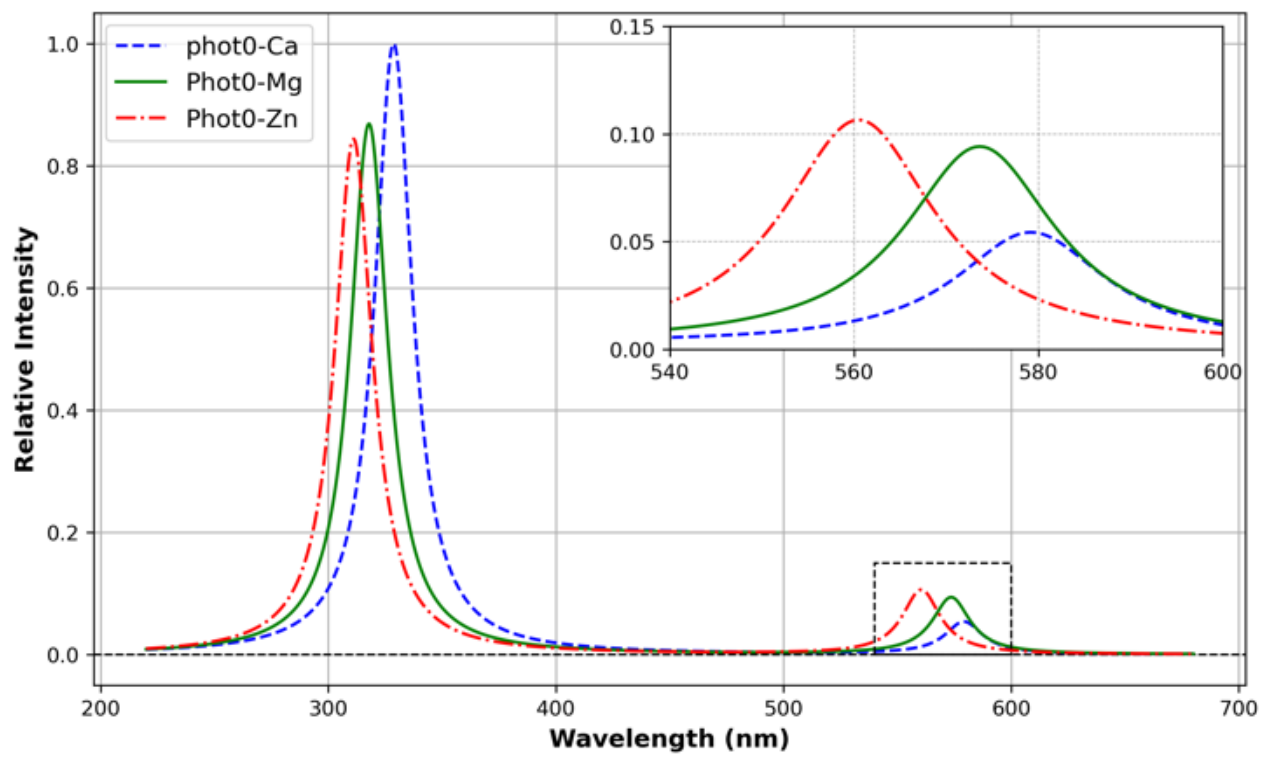

**Fig. 3:** Simulated absorption spectra in the UV-VIS range for the three phot0 molecules (phot0-Ca is the dashed blue curve, phot0-Mg is the solid green curve, phot0-Zn is the red dot-dashed curve). The spectra are obtained using Lorentzian broadening of the electronic transitions, with intensities obtained at the DLPNO-STEOM-CCSD level of theory, as also reported in Table 2. The inset shows a detail of the Q-band.

Our results validate the hypothesis that changing the central metal can modulate the absorption maxima for both the Q- and B-bands. In particular, the observed shift is consistent with the hypothesis that the metal can conjugate its $p_\pi$ orbital with the $\pi$ electrons of the HOMO of the ring (the orbital with $a_{2u}$ symmetry in phot0 and the orbital with $a_2$ symmetry in Chl *a*). In agreement with the early suggestion by Gouterman [20], a more electronegative metal (such as zinc) lowers the energy of the HOMO, resulting in blue-shifted absorption bands. In contrast, a more electropositive metal (such as calcium) raises the energy of the HOMO, resulting in red-shifted bands.

To further validate our previous TD-DFT results, we also calculated the three phot0 molecules with all the exchange–correlation functionals considered in the previous section. Calculations for the B2PLYP and ωB2PLYP functionals were performed with the ORCA 5.0.4 program, while all other functionals were calculated using Q-Chem 6.1. To analyze the results, we use the same four statistical indicators introduced in the previous section and collect them in Figure 4. All the detailed results are reported in the supporting information (Tables S3–S5). Statistical results for the three phot0 molecules show that most exchange–correlation functionals describe the Q-Band in general agreement with the accurate DLPNO-STEOM-CCSD benchmark. The situation is slightly worse for the B-band, for which local functionals largely fail. The presence of overstabilized CT ghost states is also observed for a small number of functionals, which should therefore be avoided. Generally, it seems that the phot0 molecules allow for a larger percentage of optimal exact exchange, in the range of 20-50%. As reported for Chl *a*, the PW6B95 and M06 functionals perform very well. In addition to these functionals, other high-exchange Minnesota functionals perform well for these molecules, including M06-2X, MN15, and M08-SO. Range-separated functionals are, once again, quite disappointing, even if these methods appear slightly better than in the context of Chl *a*.

**Table 2.** Vertical Excitation Energies (VEEs, in eV) associated with the Q- and B-bands of phot0-Ca, phot0-Mg, and phot0-Zn, calculated using the DLPNO-STEOM-CCSD method. The state number based on the rank of the root is reported for each state in parenthesis in the energy column.

| | $Q_y$ | | | $Q_x$ | | |
|---|---|---|---|---|---|---|
| **Metallic centre:** | **Energy (eV)** | **Wavelength (nm)** | **Oscillator Strength (arbitrary units)** | **Energy (eV)** | **Wavelength (nm)** | **Oscillator Strength (arbitrary units)** |
| Ca | 2.141 (1) | 579 | 0.063 | 2.141 (2) | 579 | 0.064 |
| Mg | 2.165 (1) | 574 | 0.112 | 2.166 (2) | 574 | 0.112 |
| Zn | 2.210 (1) | 561 | 0.127 | 2.214 (2) | 560 | 0.127 |
| | $B_x$ | | | $B_y$ | | |
| | **Energy (eV)** | **Wavelength (nm)** | **Oscillator Strength (arbitrary units)** | **Energy (eV)** | **Wavelength (nm)** | **Oscillator Strength (arbitrary units)** |
| Ca | 3.770 (5) | 329 | 1.205 | 3.775 (6) | 329 | 1.204 |
| Mg | 3.900 (4) | 318 | 1.046 | 3.900 (5) | 318 | 1.046 |
| Zn | 3.985 (4) | 311 | 1.019 | 3.986 (5) | 311 | 1.013 |

| Method: | Type | B-MUE | #GS | wMUE |
|---|---|---|---|---|
| **B2PLYP** | DH | 0.271 | 0.33 | 0.231 |
| **PW6B95** | GH (28%) | 0.135 | 0.33 | 0.153 |
| **ωB2PLYP** | RS-DH | 0.058 | 0.00 | 0.206 |
| **M06** | GH (27%) | 0.189 | 0.00 | 0.121 |
| **B3LYP** | GH (20%) | 0.188 | 0.67 | 0.201 |
| **MN15** | GH (44%) | 0.112 | 0.33 | 0.108 |
| **M08-SO** | GH (57%) | 0.103 | 0.00 | 0.106 |
| **CAM-B3LYP** | RSH (19–100%) | 0.078 | 0.33 | 0.146 |
| **PBE0** | GH (25%) | 0.108 | 0.33 | 0.160 |
| **LC-ωPBEh** | RSH (20–100%) | 0.076 | 0.67 | 0.180 |
| **ωB97X-D** | RSH (17–100%) | 0.077 | 0.67 | 0.204 |
| **LC-ωPBE** | RSH (0–100%) | 0.073 | 1.00 | 0.276 |
| **TPSSh** | GH (10%) | 0.180 | 0.67 | 0.233 |
| **M06-2X** | GH (54%) | 0.080 | 0.33 | 0.089 |
| **M08-HX** | GH (52%) | 0.082 | 0.67 | 0.173 |
| **ωM05-D** | RSH (37–100%) | 0.072 | 1.33 | 0.268 |
| **B97M-V** | L | 0.167 | 0.67 | 0.209 |
| **r2SCANh** | GH (10%) | 0.089 | 0.67 | 0.228 |
| **r2SCAN0** | GH (25%) | 0.078 | 0.33 | 0.183 |
| **ωB97X-D3** | RSH (20–100%) | 0.087 | 0.67 | 0.283 |
| **ωB97M-V** | RSH (15–100%) | 0.086 | 0.33 | 0.317 |
| **ωM06-D3** | RSH (27–100%) | 0.098 | 1.67 | 0.417 |
| **SCAN** | L | 0.147 | 0.33 | 0.130 |
| **MN15-L** | L | 0.084 | 0.33 | 0.214 |
| **M06-L** | L | 0.164 | 1.00 | 0.270 |
| **r2SCAN** | L | 0.160 | 0.67 | 0.246 |
| **revM11** | RSH (23–100%) | 0.081 | 0.67 | 0.445 |
| **ωB97X-V** | RSH (17–100%) | 0.099 | 0.67 | 0.359 |
| **TPSS** | L | 0.255 | 1.33 | 0.276 |
| **M11** | RSH (43–100%) | 0.080 | 0.33 | 0.218 |
| **BP86** | L | 0.352 | 0.67 | 0.261 |
| **PBE** | L | 0.349 | 0.33 | 0.229 |
| **BLYP** | L | 0.388 | 0.33 | 0.245 |

**Fig. 4:** Summary of the mean unsigned errors of the exchange–correlation functionals for the calculation of the absorption spectra of the three phot0 molecules calculated with respect to the reference DLPNO-STEOM-CCSD benchmark. The statistical indicators represent the following: Q-MUE: mean unsigned error for the 6 transitions of the Q-band (two transitions for each molecule); B-MUE: mean unsigned error for the 6 transitions of the B-band (two transitions for each molecule); #GS: the average number of over-stabilized CT 'ghost states' between the main signals; wMUE: the weighted MUE results calculated using Eq. 1. Each cell is shaded according to a color scale where green indicates the best performance, and red indicates the worst performance. Labels for the type of functionals are the same as in Table 1. Methods are ranked in the same order of Fig. 1.

## 4 Conclusions and Perspectives

It is well-established that the evolution of (oxygenic) photosynthesis was a transformative event in Earth's evolutionary history, and may likewise have profound consequences for extraterrestrial life on exoplanets. Chlorophylls (Chls), the primary photopigments in photosynthesis, are crucial from an astrobiological perspective because they give rise to a major biosignature known as the vegetation red edge (VRE), which corresponds to the sharp decline in absorbance (and corresponding steep increase in reflectance) of these molecules at wavelengths greater than those of the Q-band (i.e., manifested at ~700 nm).

Yet, despite the importance of photosynthesis and photopigments like Chls, the number of studies that have sought to model the structures and absorption spectra of the latter from a computational chemistry perspective are surprisingly few in number. In this work, we have sought to rectify this issue by first benchmarking time-dependent density-functional theory (TD-DFT) absorption spectra of Chl *a*, and then using the insights from this section to investigate an alternative photopigment known as phot0.

In the case of Chl *a*, we demonstrate that two exchange–correlation functional approximations outperform the others in terms of their accuracy of rendering absorption bands and minimizing the number of so-called 'ghost states'. The first is the B2PLYP double-hybrid functional, which was already posited by previous studies. The second is the PW6B95 global-hybrid functional, which can provide an accuracy similar to B2PLYP at a more convenient computational cost and with a simpler implementation (several quantum chemistry codes lack the ability of running TD-DFT calculations with double-hybrid functionals, while global-hybrids are generally standard). General trends among TD-DFT calculations show that local functional suffer from overstabilization of charge-transfer ghost states, while range-separated functionals offer degraded performance due to red-shifting of the energy states. The optimal compromise is achieved with global-hybrid functionals with a percentage of exact exchange of around 25%. We also study the effects of certain changes in the molecular geometry (qualitative trends appear to be preserved) and the Tamm-Dancoff approximation (which appears to affect the results significantly, and is therefore not recommended).

Moving on to phot0, we explored three different configurations, where the central metal ion is either Mg, Ca, or Zn. Using calculations with the accurate DLPNO-STEOM-CCSD wave function method, we find that the electronegativity of the metal ion has a distinct effect on the absorption spectra, with more electronegative metals causing blue-shifting and the converse applies for electropositive metals. In addition, we found that the aforementioned PW6B95 and M06 functionals are effective when compared against the DLPNO-STEOM-CCSD results.

If the same trend is applicable to extraterrestrial photopigments, some interesting consequences would ensue. As mentioned in Section 2.2, it has been theorized that the absorption spectra (specifically the Q-band) of photopigments may be approximatively governed by the peak spectral photon flux of the host star. In case this hypothesis is correct, photopigments on exoplanets around hotter and cooler stars (with respect to the Sun) could be blue-shifted and red-shifted, respectively, when compared to their analogs on Earth. This blue-shifting and red-shifting might, in turn, be facilitated by the evolution of photopigments comprising metals with higher and lower electronegativity, respectively.

This work represents a preliminary study, and further research is needed to optimize the performance of computational methods for prediction of absorption spectra of photopigments in astrobiological conditions. Particularly, it is essential to explore the intricacies of predicting absorption spectra, encompassing aspects like understanding the significance of vibronic structure, the influence of temperature, the effects of the environment, forecasting fluorescence and phosphorescence rates, and so forth. By doing so, we may arrive at a deeper understanding of the absorption properties of putative extraterrestrial photopigments, which in turn can enable us to seek them out with future generations of ground- and space-based telescopes.

## Dedication

This work is dedicated to Dr. Timothy J. Lee, pioneer in the field of astrochemistry, mentor and collaborator of R.P. for several years. As we dedicate this article to Dr. Lee's memory, we do so with a heavy heart, knowing that he has left us too soon. Yet, we also celebrate his contributions to the scientific communities of computational chemistry and astrochemistry. Dr Lee's commitment to fostering scientific curiosity and nurturing intellectual growth left an enduring impact on countless lives, including R.P.'s. His legacy lives on through the knowledge he imparted and the lessons he instilled. So long Dr. Lee, and thanks for all the fish.

## Acknowledgements

D.I. and R.P. would like to thank Pierpaolo Morgante for useful suggestions and discussions. M.L. would like to thank Boris Akhremitchev for stimulating discussions of the nature of putative extraterrestrial photopigments.

*Supporting Information for*

# On the importance and challenges of modelling extraterrestrial photopigments via density-functional theory

**Dorothea Illner,[1] Manasvi Lingam,[2] and Roberto Peverati[1*]**

[1] Department of Chemistry and Chemical Engineering, Florida Institute of Technology, Melbourne, FL - 32901, USA
[2] Department of Aerospace, Physics, and Space Sciences, Florida Institute of Technology, Melbourne, FL - 32901, USA

Email: rpeverati@fit.edu

**Geometry S1: Geometry of the smallest significantly representative Chlorophyll *a* molecule in gas phase (xyz format) optimized using the CAM-B3LYP-D3(BJ)/def2-TZVP method in gas phase.**

```
76

C   -6.94101773277842    35.79286800103698    7.98635477633448
C   -3.98102222294773    34.72352147911032   -0.68162218643282
C    2.41818178868599    37.10574470780596    3.41245195749761
C   -2.25219108797725    35.83410795974775    9.92666317325171
N   -4.55573995402007    34.85244169287340    5.83191257776971
H   -6.95620340674774    36.52504882196585    7.17750760283703
C   -4.89481255725964    34.97140548072344   -1.61150116602348
C    2.48181598256915    38.63053884709370    3.49611396655616
C   -3.73210795994643    35.42402953241709    9.54429168960854
N   -3.75420079300643    34.95221437559223    3.00064582029123
O   -1.87336348390208    35.97437278633750    11.0593395452471
C   -4.09749968731009    34.15934041884622    10.2785200864900
N   -1.85942340124899    35.79350469942186    6.43005581516318
C   -4.92442673655606    33.30847807206517    12.2946151224424
C   -3.72674232147732    35.30841136162393    8.03371472034305
C   -5.87762230053422    34.17910143776908    3.90490125603511
C   -1.94637567123760    35.53608042684566    1.47232676040621
C    0.33822428813312    36.44395180455594    5.64154956037812
C   -6.70801062326837    32.37717845011613    6.39738799170638
C   -6.58976128707013    33.75127458331651    0.91611699647059
C    0.90309353380740    36.48799076888075    0.65901366786070
C    0.92850593286370    36.72375011779826    8.76209709493966
C   -4.72304533128727    34.92584886159361    7.19675013858754
C   -5.00591773439092    34.47481367927818    2.85622278121490
C   -0.94550997379709    35.91021056990892    2.39554702753548
C   -0.55645851146418    36.20347206271686    6.67434590787102
O   -3.90057604920894    33.04441109251169    9.87900551397264
C   -6.13513701434933    34.56037242607884    7.58297763862106
C   -5.31260729403626    34.30217757237480    1.44501939468094
C    0.37992399416866    36.34352905300695    2.04797417964627
C   -0.33715210100630    36.32969602849064    8.08673771856715
O   -4.65530300455787    34.43366942549807    11.4594334667058
C   -6.66247388997242    33.90010713763338    6.30004821256342
C   -4.20805986972384    34.69978888605092    0.75626458681333
C    1.02818987790773    36.59274924832584    3.22622687259362
C   -1.54709271266898    35.98852019415750    8.66566409034399
C   -5.65579855871918    34.34437824565367    5.25983506310835
C   -3.21234644663621    35.09577835898741    1.74408156169555
C    0.08919240891217    36.31548071091521    4.28120403180639
C   -2.43004620330514    35.67266987424515    7.61435309236558
N   -1.10207774797484    35.90248175662032    3.73330560569733
Mg  -2.81936958416003    35.37504280006335    4.73082480926422
H   -6.51971340550882    36.27488130000832    8.86751455378494
H   -7.97120444011421    35.51679715844789    8.21470700459646
H   -2.96357020588187    34.52917560886468   -1.00361483443642
H    3.04461322977542    36.76349709668155    2.58723086727845
H    2.85418994697830    36.67397392994451    4.31513058727761
H   -5.91807998671285    35.21539620900278   -1.36226154263513
H   -4.63548724889499    34.96147919848033   -2.66111134393964
H    3.50762470000446    38.97339232905264    3.63593605015654
H    1.88248030415175    39.00046835447498    4.32836403859072
```

```
H    2.09254106863622    39.08456431663874    2.58446896614781
H   -4.37267789836554    36.22623887624258    9.91291470387167
H   -5.35285359181750    33.71337477307435    13.2058430354878
H   -4.00301089535288    32.77154508885973    12.5112276526719
H   -5.62426234218440    32.63084292891680    11.8087020245978
H   -6.84118958385343    33.77352009924613    3.63134868089604
H   -1.67224832888263    35.60406296471633    0.42903692747055
H    1.33111463989666    36.76910223600049    5.92329432387634
H   -7.00847192519247    31.92856356107021    5.45099676325499
H   -7.41832223918920    32.06612952140186    7.16412981773137
H   -5.72744775981967    31.98137481182714    6.66516706338777
H   -6.44267602504536    33.31291045675245   -0.06958849406966
H   -7.35614826273716    34.52438689221510    0.82150028117975
H   -6.98788109586424    32.97978677534699    1.57448868840590
H    1.95918844675213    36.75121695075061    0.66129703343982
H    0.37308342580854    37.26896561054314    0.10979316327020
H    0.79621656376052    35.56243883975556    0.09051666038683
H    1.36208053891397    37.61728465720379    8.31128588176325
H    1.67293763434527    35.92795432952511    8.69090639248266
H    0.75143746270652    36.91972204710151    9.81717960470833
H   -6.12906693109937    33.84721960099380    8.40808224216130
H   -7.65692130171338    34.27235505080080    6.04711242237590
```

**Geometry S2: Geometry of the full Chlorophyll *a* molecule in gas phase (xyz format) optimized using the GFN2-xtb method.**

```
135
 energy: -186.255236093884 gnorm: 0.000859732492 xtb: 6.4.1 (conda-forge)
Mg       -1.16538284212160   -3.52620922994169    1.64790409848501
C        -1.72747203367702   -1.76761321193830   -1.23089086841779
C         1.66823597014874   -4.75838167490581    0.32994391849361
C        -0.69343026900199   -5.27254738249884    4.49206996217744
C        -4.04656394429298   -2.14479006977554    3.00539804209098
N        -0.19662809723576   -3.26024453352374   -0.16775000259207
C        -0.53459828725249   -2.47769116448398   -1.23316646063359
C         0.50464868809369   -2.54081589924262   -2.23989645923886
C         1.44720727211531   -3.40349524867605   -1.76318239779175
C         0.98229966155329   -3.85220937796596   -0.46634557168418
C         2.72896369261202   -3.82425815897218   -2.39309475774211
C         0.57365618479885   -1.69564539774623   -3.46509324333628
C         1.00262146723003   -0.26751054526414   -3.07937739703660
C         0.55569952357929    0.74129261705181   -4.11312785264377
O        -0.55720433453401    0.80369534086824   -4.56456536939466
O         1.55274310076092    1.56118940843654   -4.45281406870765
N         0.22891277564113   -4.79224490998077    2.29414966524913
C         1.31928996499788   -5.19690301226265    1.59825819692085
C         2.07761592406553   -6.14986619749014    2.37703056010406
C         1.40674419191022   -6.30128655562902    3.56342193352044
C         0.25044985757622   -5.42436218434864    3.49030017253021
C         3.36218431652437   -6.77328783540607    1.95948671886414
C         1.74657646957665   -7.12665213698784    4.70363157834626
C         2.31719057238856   -8.32440597182702    4.63931194680009
N        -2.18790091413070   -3.67038409014718    3.42416483839362
C        -1.81264887425836   -4.45080785969284    4.46233452820020
C        -2.75034708445467   -4.30921830091531    5.55772551803540
```

| | | | |
|---|---|---|---|
| C | -3.70033417664971 | -3.42623768794982 | 5.13907605576302 |
| C | -3.32967395772681 | -3.03142070688373 | 3.79455785585312 |
| C | -2.63829832281422 | -5.01244777216940 | 6.86538150789951 |
| C | -4.87425662223477 | -2.90058883384174 | 5.89600878834615 |
| C | -4.54772359464972 | -1.55237984404117 | 6.54348659512401 |
| N | -2.57469049661477 | -2.24313951868939 | 1.09697994624268 |
| C | -3.69798904659966 | -1.77315954046770 | 1.71361944183414 |
| C | -4.43403074305398 | -0.88665730390507 | 0.83521263709957 |
| C | -3.70864456614667 | -0.86802679078098 | -0.33493256406632 |
| C | -2.57435507686125 | -1.71829795354830 | -0.12786346089980 |
| C | -5.69578003415131 | -0.17396737949443 | 1.15160418829544 |
| C | -3.65300942228294 | -0.32463071575385 | -1.67718010526978 |
| O | -4.36567053543620 | 0.48268628881820 | -2.22025254310386 |
| C | -2.43439194934823 | -1.01991459599886 | -2.34886249930483 |
| C | -2.97472827840259 | -2.08385291892413 | -3.29501146514508 |
| O | -2.52268584253199 | -2.35907681942995 | -4.37233365150161 |
| O | -4.01132304661814 | -2.72106413661209 | -2.74023909850497 |
| C | -4.58958000479275 | -3.76186444804531 | -3.51401789019261 |
| C | 1.24782376008659 | 2.59414388852101 | -5.38909383407633 |
| C | 0.48145382005251 | 3.71902355388093 | -4.75883025489930 |
| C | 0.29017166033831 | 4.90428043595292 | -5.32456471591100 |
| C | 0.86859614831605 | 5.24194901473929 | -6.66831360829723 |
| C | -0.50412731475091 | 5.97888893015590 | -4.63988414370957 |
| C | 0.38610415152231 | 7.12354700212461 | -4.13910424396372 |
| C | 1.37373468013415 | 6.64577645411978 | -3.07721741727045 |
| C | 2.11294482513197 | 7.78306640627882 | -2.36134671710124 |
| C | 2.99138624186205 | 8.56957236162826 | -3.33143328376083 |
| C | 2.94695898429992 | 7.20499246978867 | -1.20981594180794 |
| C | 3.44163579111084 | 8.26166944089888 | -0.22482691603311 |
| C | 4.08371041453149 | 7.67301846620609 | 1.03373593255404 |
| C | 5.41362375007701 | 6.93996902071311 | 0.77783705793620 |
| C | 5.22680061845921 | 5.42450291280797 | 0.71773760961953 |
| C | 6.47473112884480 | 7.32244795251007 | 1.82146086827407 |
| C | 6.14917147542171 | 6.87395974523293 | 3.24612492810344 |
| C | 7.08178770231106 | 7.50435356210708 | 4.28363119395853 |
| C | 6.77823780994297 | 8.97347844650776 | 4.60391337986947 |
| C | 7.95557005773624 | 9.59760830367302 | 5.35127854610959 |
| C | 5.50636773774435 | 9.11244565232631 | 5.43768415560753 |
| H | 2.57730431684439 | -5.16678034767995 | -0.08797496562138 |
| H | -0.54447070239356 | -5.84963909003847 | 5.39351278105665 |
| H | -4.95017093754695 | -1.71992855643455 | 3.41763650754448 |
| H | 2.84607719095787 | -3.36924324834704 | -3.37269628948947 |
| H | 2.75964199393159 | -4.90725258352245 | -2.51098058726933 |
| H | 3.57413584471415 | -3.52682265794763 | -1.77220167896766 |
| H | -0.39295983118390 | -1.66397396059429 | -3.96714163424154 |
| H | 1.29621466193653 | -2.11132356066069 | -4.16872606616872 |
| H | 2.08196297991798 | -0.19938490485613 | -2.94564736863400 |
| H | 0.52213469236851 | 0.00867124970544 | -2.13618990605550 |
| H | 3.98077940380148 | -6.05785078931455 | 1.42039784558446 |
| H | 3.18200383653617 | -7.62581281428438 | 1.30297132153973 |
| H | 3.90988561613889 | -7.12161509904052 | 2.83255837763350 |
| H | 1.49614251941906 | -6.71096502106360 | 5.67234911350557 |
| H | 2.54204883273672 | -8.80215579773017 | 3.70180493380924 |
| H | 2.55405900061371 | -8.88306410349887 | 5.52906390016039 |
| H | -2.60969242227618 | -6.09216670398333 | 6.71927495069210 |
| H | -3.48228053635902 | -4.77238061253773 | 7.50662060155305 |
| H | -1.72316212598440 | -4.71651904528658 | 7.37872971256396 |
| H | -5.15962963461987 | -3.61275896954314 | 6.67287265982204 |
| H | -5.72782695135396 | -2.78517911798576 | 5.22462834048553 |
| H | -3.72206199531607 | -1.66167492473800 | 7.24291786170139 |
| H | -5.41145923755952 | -1.16966352133424 | 7.08280087410791 |
| H | -4.25834778885899 | -0.82700711184577 | 5.78655382162695 |
| H | -6.00208094254414 | 0.42625460805515 | 0.29902310589007 |
| H | -5.56120412504182 | 0.47759224434896 | 2.01529338462712 |
| H | -6.48770323818968 | -0.88487030972681 | 1.38826492198943 |
| H | -1.83301863545756 | -0.31553720825276 | -2.92600100554937 |
| H | -5.42342709913094 | -4.14485215260154 | -2.93167003910340 |
| H | -3.85472231388739 | -4.54855934961014 | -3.69800836301996 |
| H | -4.93662620713719 | -3.37124393296215 | -4.47261768584195 |
| H | 0.66607970313461 | 2.16250363951622 | -6.21544103603206 |
| H | 2.21693373657634 | 2.92946703860355 | -5.76871087435386 |
| H | 0.04998941541584 | 3.49415806958468 | -3.79478465954370 |
| H | 0.59306420758646 | 4.48720118900350 | -7.40323674563998 |
| H | 1.95732319953208 | 5.27388632419904 | -6.62112021861756 |
| H | 0.51116960191050 | 6.20797505945417 | -7.01373005179215 |
| H | -1.04775795518093 | 5.55266020288102 | -3.79542493923799 |
| H | -1.23653778402546 | 6.38281798759584 | -5.34361257269761 |
| H | 0.92430098079735 | 7.55507013501816 | -4.98341956300923 |
| H | -0.25174185754014 | 7.90168369019441 | -3.71325327349207 |
| H | 2.10155942290351 | 5.97300872104440 | -3.53558396893542 |
| H | 0.82123794987988 | 6.06862206424634 | -2.33169707335354 |
| H | 1.36503643580538 | 8.46398962450708 | -1.93777986864333 |
| H | 3.71091292735257 | 7.90793774835170 | -3.80986097617300 |
| H | 3.53646304316697 | 9.35161319466102 | -2.80970222792918 |
| H | 2.39024318747629 | 9.04268064869608 | -4.10310009190475 |
| H | 3.79604526995120 | 6.66457550281316 | -1.63126003257985 |
| H | 2.33304858666801 | 6.48704222125054 | -0.66020913644676 |
| H | 2.59063155160076 | 8.87707134507802 | 0.08019664442559 |
| H | 4.16453150466074 | 8.91911747044809 | -0.71147114359464 |
| H | 3.37390642882763 | 6.99475912813061 | 1.51158968831841 |
| H | 4.26135008348222 | 8.50007970506054 | 1.72494549776886 |
| H | 5.80035817953021 | 7.26953338043653 | -0.19376230184324 |
| H | 4.79005105292772 | 5.05226868379009 | 1.64098198165494 |
| H | 6.18525524856110 | 4.93125478944869 | 0.56701142558433 |
| H | 4.56865422340969 | 5.14798507606799 | -0.10115284169874 |
| H | 6.59374285634050 | 8.40786700765699 | 1.80118380333329 |
| H | 7.43362365154270 | 6.88716237702010 | 1.52727887849925 |
| H | 6.24950635266939 | 5.78913956891062 | 3.30868765889223 |
| H | 5.11507289427569 | 7.12233529345115 | 3.48395329303705 |
| H | 8.10861416698437 | 7.42555755218166 | 3.91756214354397 |
| H | 7.02163054192390 | 6.93244731691120 | 5.21296050640836 |
| H | 6.63940952322834 | 9.52192169994165 | 3.66634630440240 |
| H | 8.14057352049245 | 9.06148835005191 | 6.28008089108058 |
| H | 7.74887864196941 | 10.63817199985170 | 5.59205678499662 |
| H | 8.85979209923802 | 9.55943246910748 | 4.74758732137893 |
| H | 5.61732909556768 | 8.58987354540202 | 6.38589144852195 |
| H | 4.64606894214337 | 8.69917716653475 | 4.91752047758773 |
| H | 5.30252070741565 | 10.16062050275699 | 5.64631637113647 |

**Geometry S3: Geometry of the phot0-Ca molecule in gas phase (xyz format) optimized using the CAM-B3LYP-D3(BJ)/def2-TZVP method.**

```
29

C       2.9892391854     1.1471639091     0.0034393674
C       2.4638805878     2.4381330409    -0.0039694790
C       1.1784519546     2.9769331795    -0.0085584206
N       2.3225414601    -0.0121581292     0.0042383118
N       0.0121715724     2.3224880802    -0.0042785723
C      -1.1471309872     2.9892354245    -0.0033572678
C      -2.4381007431     2.4639293245     0.0041500574
C      -2.9769261843     1.1784963723     0.0086679069
N      -2.3224946786     0.0122195575     0.0043313296
C      -2.9892139693    -1.1471036006     0.0031012194
C      -2.4639065337    -2.4380872921    -0.0039924460
C      -1.1784984337    -2.9769717240    -0.0082368063
N      -0.0122068398    -2.3225499012    -0.0042640142
C       1.1471368247    -2.9892192361    -0.0034051080
C       2.4381161307    -2.4638588263     0.0039778354
C       2.9769730229    -1.1784536098     0.0085804375
Ca      0.0000278399    -0.0000698166    -0.0001535426
H       4.0779258839     1.0828995018     0.0075183379
H       3.2335695403     3.2000345355    -0.0059616869
H       1.1255994786     4.0662292161    -0.0138300289
H      -1.0828199680     4.0779191222    -0.0074187649
H      -3.2000061817     3.2336140332     0.0061883761
H      -4.0662232327     1.1256598953     0.0140090715
H      -4.0779000231    -1.0828074929     0.0066925340
H      -3.2337327048    -3.1998507829    -0.0060662679
H      -1.1256935708    -4.0662715091    -0.0131424936
H       1.0828895441    -4.0779072086    -0.0073630619
H       3.1998955002    -3.2336694883     0.0060703889
H       4.0662689040    -1.1256331621     0.0140360110
```

**Geometry S4: Geometry of the phot0-Mg molecule in gas phase (xyz format) optimized using the CAM-B3LYP-D3(BJ)/def2-TZVP method.**

```
29

C       1.3689887787    -2.7790501944    -0.0000000000
C       0.1543492219    -3.4222136696     0.0000000000
C      -1.1132294773    -2.8910067338     0.0000000000
N       1.5922523169    -1.4546031633    -0.0000000000
N      -1.4545681959    -1.5921736076    -0.0000000000
C      -2.7790183151    -1.3689094887    -0.0000000000
C      -3.4222043245    -0.1542819408    -0.0000000000
C      -2.8910235135     1.1133080687    -0.0000000000
N      -1.5921961557     1.4546588145    -0.0000000000
C      -1.3689312073     2.7791053598    -0.0000000000
C      -0.1542936558     3.4222745409     0.0000000000
C       1.1132878784     2.8910706284     0.0000000000
N       1.4546301085     1.5922401790    -0.0000000000
C       2.7790796708     1.3689741471    -0.0000000000
C       3.4222600660     0.1543436440    -0.0000000000
C       2.8910775150    -1.1132472011    -0.0000000000
```

| | | | |
|---|---|---|---|
| Mg | 0.0000288916 | 0.0000060092 | -0.0000000000 |
| H | 2.2527652200 | -3.4128227569 | 0.0000000000 |
| H | 0.2028120807 | -4.5028261269 | 0.0000000000 |
| H | -1.9366164862 | -3.6014601610 | 0.0000000000 |
| H | -3.4127849601 | -2.2526901033 | -0.0000000000 |
| H | -4.5028155388 | -0.2027712393 | 0.0000000000 |
| H | -3.6014901216 | 1.9366847258 | 0.0000000000 |
| H | -2.2527070310 | 3.4128801907 | -0.0000000000 |
| H | -0.2027619164 | 4.5028873142 | 0.0000000000 |
| H | 1.9366720074 | 3.6015290174 | 0.0000000000 |
| H | 3.4128508111 | 2.2527524639 | 0.0000000000 |
| H | 4.5028720094 | 0.2028271116 | -0.0000000000 |
| H | 3.6015469516 | -1.9366217292 | -0.0000000000 |

**Geometry S5: Geometry of the phot0-Zn molecule in gas phase (xyz format) optimized using the CAM-B3LYP-D3(BJ)/def2-TZVP method.**

29

| | | | |
|---|---|---|---|
| C | -1.6758236900 | -2.5900315753 | -0.0128403480 |
| C | -2.8534355177 | -1.8881045675 | -0.0036397532 |
| C | -3.0408011703 | -0.5301473973 | 0.0022005989 |
| N | -0.4279085833 | -2.0950663138 | -0.0135152410 |
| N | -2.0985791303 | 0.4261313548 | -0.0025364209 |
| C | -2.5934764191 | 1.6739962782 | -0.0029149319 |
| C | -1.8916345680 | 2.8515611934 | -0.0125767263 |
| C | -0.5336833615 | 3.0389883937 | -0.0187279729 |
| N | 0.4225862473 | 2.0968359543 | -0.0136679199 |
| C | 1.6705293845 | 2.5918616972 | -0.0121459653 |
| C | 2.8481617450 | 1.8900020483 | -0.0033911578 |
| C | 3.0355801777 | 0.5319451537 | 0.0019916388 |
| N | 2.0932300764 | -0.4243456774 | -0.0023616597 |
| C | 2.5883474994 | -1.6723540147 | -0.0025842168 |
| C | 1.8864521254 | -2.8500491291 | -0.0120451513 |
| C | 0.5283980201 | -3.0374461632 | -0.0184697843 |
| Zn | -0.0028526776 | 0.0010467845 | -0.0077544672 |
| H | -1.7546913347 | -3.6741123238 | -0.0180097634 |
| H | -3.7548130113 | -2.4856532124 | -0.0013921247 |
| H | -4.0702412088 | -0.1811624795 | 0.0083601292 |
| H | -3.6774447541 | 1.7528563203 | 0.0028937675 |
| H | -2.4891254108 | 3.7529227486 | -0.0147183217 |
| H | -0.1847046532 | 4.0684423686 | -0.0253096347 |
| H | 1.7493946023 | 3.6759035960 | -0.0164511014 |
| H | 3.7496318255 | 2.4874993177 | -0.0010847786 |
| H | 4.0651479737 | 0.1828729863 | 0.0076692170 |
| H | 3.6725010435 | -1.7511845357 | 0.0030350451 |
| H | 2.4840548017 | -3.7515038392 | -0.0140861413 |
| H | 0.1793697294 | -4.0670019633 | -0.0252220694 |

**Table S1:** Vertical Excitation Energies (VEEs, in eV) associated with the Q- and B-bands of Chlorophyll *a* calculated using DLPNO-STEOM-CCSD and TD-DFT with various exchange–correlation functionals with geometry S2.

| **Method:** | **$Q_y$** | **State #** | **$Q_x$** | **State #** | **$B_x$** | **State #** | **$B_y$** | **State #** |
|---|---|---|---|---|---|---|---|---|
| **Exp** | **1.99** | **1** | **2.3** | **2** | **3.12** | **3** | **3.38** | **4** |
| **DLPNO** | **1.75** | **1** | **2.24** | **2** | **3.17** | **3** | **3.40** | **4** |
| BLYP | 2.06 | 1 | 2.09 | 2 | 3.06 | 11 | 3.19 | 16 |
| PBE | 2.07 | 1 | 2.10 | 2 | 2.90 | 11 | 3.10 | 15 |
| BP86 | 2.07 | 1 | 2.10 | 2 | 2.84 | 12 | 3.08 | 15 |
| M06-L | 2.16 | 1 | 2.17 | 2 | 3.02 | 11 | 3.30 | 14 |
| MN15-L | 2.29 | 1 | 2.31 | 2 | 3.10 | 11 | 3.22 | 14 |
| TPSS | 2.11 | 1 | 2.14 | 2 | 2.94 | 11 | 3.16 | 15 |
| SCAN | 2.16 | 1 | 2.17 | 2 | 3.00 | 11 | 3.11 | 15 |
| B97M-V | 2.19 | 1 | 2.21 | 2 | 2.99 | 11 | 3.12 | 15 |
| B3LYP | 2.22 | 1 | 2.26 | 2 | 3.13 | 3 | 3.21 | 4 |
| PBE0 | 2.25 | 1 | 2.29 | 2 | 3.15 | 3 | 3.23 | 4 |
| TPSSh | 2.21 | 1 | 2.22 | 2 | 3.10 | 3 | 3.18 | 4 |
| r2SCANh | 2.26 | 1 | 2.28 | 2 | 3.20 | 3 | 3.29 | 4 |
| r2SCAN0 | 2.30 | 1 | 2.35 | 2 | 3.25 | 3 | 3.33 | 4 |
| PW6B95 | 2.24 | 1 | 2.28 | 2 | 3.15 | 3 | 3.23 | 4 |
| M06 | 2.16 | 1 | 2.21 | 2 | 3.12 | 3 | 3.20 | 4 |
| M06-2X | 2.25 | 1 | 2.32 | 2 | 3.36 | 3 | 3.41 | 4 |
| M08-HX | 2.28 | 1 | 2.35 | 2 | 3.36 | 3 | 3.41 | 4 |
| MN15 | 2.22 | 1 | 2.28 | 2 | 3.25 | 3 | 3.31 | 4 |
| CAM-B3LYP | 2.17 | 1 | 2.24 | 2 | 3.35 | 3 | 3.40 | 4 |
| ωB97M-V | 2.02 | 1 | 2.09 | 2 | 3.40 | 3 | 3.43 | 4 |
| ωB97X-V | 2.01 | 1 | 2.09 | 2 | 3.45 | 3 | 3.48 | 4 |
| M11 | 2.08 | 1 | 2.16 | 2 | 3.42 | 3 | 3.45 | 4 |
| revM11 | 1.91 | 1 | 2.00 | 2 | 3.42 | 3 | 3.43 | 4 |

**Table S2:** Vertical Excitation Energies (VEEs, in eV) associated with the Q- and B-bands of Chlorophyll *a* calculated using regular TD-DFT (left column, same values reported in the main text) and TD-DFT with the Tamm–Dancoff approximation (TDA, right column) with various exchange–correlation functionals with geometry S2. The last two columns report the mean unsigned errors (MUE) of both methods.

| | Qy | | Qx | | Bx | | By | | | |
|---|---|---|---|---|---|---|---|---|---|---|
| **Exp** | **1.99** | | **2.30** | | **3.12** | | **3.38** | | | |
| | **TDDFT** | **TDA** | **TDDFT** | **TDA** | **TDDFT** | **TDA** | **TDDFT** | **TDA** | **MUE TDDFT** | **MUE TDA** |
| BLYP | 2.05 | 2.13 | 2.13 | 2.14 | 2.82 | 2.86 | 2.95 | 2.96 | 0.24 | 0.25 |
| PBE | 2.07 | 2.13 | 2.14 | 2.15 | 2.83 | 2.85 | 2.96 | 2.98 | 0.24 | 0.24 |
| BP86 | 2.06 | 2.15 | 2.14 | 2.36 | 2.84 | 3.01 | 2.96 | 3.11 | 0.23 | 0.15 |
| M06-L | 2.13 | 2.27 | 2.24 | 2.33 | 3.01 | 3.18 | 3.15 | 3.27 | 0.14 | 0.12 |
| MN15-L | 2.23 | 2.38 | 2.38 | 2.45 | 3.17 | 3.36 | 3.34 | 3.57 | 0.10 | 0.24 |
| r2SCAN | 2.16 | 2.25 | 2.26 | 2.26 | 3.01 | 3.06 | 3.15 | 3.21 | 0.14 | 0.13 |
| B97M-V | 2.15 | 2.25 | 2.28 | 2.26 | 3.05 | 3.12 | 3.20 | 3.25 | 0.11 | 0.11 |
| B3LYP | 2.16 | 2.31 | 2.35 | 2.43 | 3.17 | 3.13 | 3.35 | 3.42 | 0.07 | 0.13 |
| r2SCANh | 2.20 | 2.32 | 2.35 | 2.33 | 3.17 | 3.32 | 3.33 | 3.38 | 0.09 | 0.14 |
| r2SCAN0 | 2.24 | 2.38 | 2.47 | 2.41 | 3.36 | 3.43 | 3.56 | 3.64 | 0.21 | 0.27 |
| PW6B95 | 2.17 | 2.31 | 2.40 | 2.35 | 3.22 | 3.30 | 3.27 | 3.35 | 0.12 | 0.15 |
| M06 | 2.11 | 2.25 | 2.35 | 2.28 | 3.23 | 3.33 | 3.42 | 3.51 | 0.08 | 0.16 |
| M06-2X | 2.19 | 2.39 | 2.56 | 2.45 | 3.44 | 3.77 | 3.69 | 3.87 | 0.27 | 0.42 |
| M08-HX | 2.21 | 2.4 | 2.56 | 2.68 | 3.44 | 3.69 | 3.68 | 3.8 | 0.27 | 0.45 |
| M08-SO | 2.15 | 2.34 | 2.54 | 2.66 | 3.41 | 3.71 | 3.68 | 3.8 | 0.25 | 0.43 |
| MN15 | 2.15 | 2.33 | 2.47 | 2.38 | 3.33 | 3.58 | 3.56 | 3.71 | 0.18 | 0.30 |
| LC-wPBE | 2.07 | 2.26 | 2.59 | 2.29 | 3.47 | 3.37 | 3.78 | 3.45 | 0.28 | 0.15 |
| CAM-B3LYP | 2.14 | 2.33 | 2.53 | 2.39 | 3.43 | 3.76 | 3.71 | 3.85 | 0.26 | 0.39 |

**Table S3:** Vertical Excitation Energies (VEEs, in eV) associated with the Q- and B-bands of phot0-Ca calculated using DLPNO-STEOM-CCSD and TD-DFT with various exchange–correlation functionals with the geometry in S3.

| | $Q_y$ | $Q_x$ | $B_y$ | $B_x$ |
|---|---|---|---|---|
| **DLPNO-STEOM-CCSD** | 2.141 | 2.141 | 3.77 | 3.775 |
| **B2PLYP** | 2.219 | 2.219 | 3.421 | 3.421 |
| **ωB2PLYP** | 1.971 | 1.971 | 3.615 | 3.615 |
| **PW6B95** | 2.195 | 2.195 | 3.513 | 3.513 |
| **M06** | 2.116 | 2.116 | 3.468 | 3.468 |
| **B3LYP** | 2.174 | 2.174 | 3.448 | 3.448 |
| **MN15** | 2.135 | 2.135 | 3.538 | 3.538 |
| **M08-SO** | 2.099 | 2.099 | 3.557 | 3.447 |
| **CAM-B3LYP** | 2.069 | 2.069 | 3.592 | 3.592 |
| **PBE0** | 2.219 | 2.219 | 3.547 | 3.547 |
| **LC-ωPBEh** | 2.067 | 2.067 | 3.626 | 3.626 |
| **ωB97X-D** | 2.043 | 2.043 | 3.623 | 3.623 |
| **LC-ωPBE** | 1.968 | 1.968 | 3.635 | 3.635 |
| **TPSSh** | 2.213 | 2.213 | 3.477 | 3.477 |
| **M06-2X** | 2.165 | 2.165 | 3.616 | 3.616 |
| **M08-HX** | 2.216 | 2.216 | 3.575 | 3.624 |
| **ωM05-D** | 2.053 | 2.053 | 3.647 | 3.647 |
| **B97M-V** | 2.180 | 2.180 | 3.458 | 3.458 |
| **$r^2$SCANh** | 2.258 | 2.258 | 3.567 | 3.567 |
| **$r^2$SCAN0** | 2.263 | 2.263 | 3.664 | 3.664 |
| **ωB97X-D3** | 1.969 | 1.969 | 3.659 | 3.659 |
| **ωB97M-V** | 1.894 | 1.894 | 3.597 | 3.597 |
| **ωM06-D3** | 1.940 | 1.940 | 3.670 | 3.670 |
| **SCAN** | 2.110 | 2.110 | 3.501 | 3.501 |
| **MN15-L** | 2.273 | 2.273 | 3.608 | 3.608 |
| **M06-L** | 2.222 | 2.222 | 3.510 | 3.510 |
| **$r^2$SCAN** | 2.230 | 2.230 | 3.492 | 3.492 |
| **revM11** | 1.796 | 1.796 | 3.660 | 3.660 |
| **ωB97X-V** | 1.895 | 1.895 | 3.679 | 3.679 |
| **TPSS** | 2.112 | 2.112 | 3.531 | 3.531 |
| **M11** | 2.006 | 2.006 | 3.681 | 3.681 |
| **BP86** | 2.129 | 2.129 | 3.284 | 3.284 |
| **PBE** | 2.133 | 2.133 | 3.287 | 3.287 |
| **BLYP** | 2.093 | 2.093 | 3.229 | 3.229 |

**Table S4:** Vertical Excitation Energies (VEEs, in eV) associated with the Q- and B-bands of phot0-Mg calculated using DLPNO-STEOM-CCSD and TD-DFT with various exchange–correlation functionals with the geometry in S4.

| | $Q_y$ | $Q_x$ | $B_y$ | $B_x$ |
|---|---|---|---|---|
| **DLPNO-STEOM-CCSD** | 2.165 | 2.166 | 3.900 | 3.900 |
| **B2PLYP** | 2.219 | 2.219 | 3.678 | 3.678 |
| **ωB2PLYP** | 1.978 | 1.978 | 3.894 | 3.894 |
| **PW6B95** | 2.216 | 2.216 | 3.835 | 3.835 |
| **M06** | 2.136 | 2.136 | 3.779 | 3.779 |
| **B3LYP** | 2.209 | 2.209 | 3.791 | 3.791 |
| **MN15** | 2.139 | 2.139 | 3.851 | 3.851 |
| **M08-SO** | 2.103 | 2.103 | 3.882 | 3.882 |
| **CAM-B3LYP** | 2.089 | 2.089 | 3.931 | 3.931 |
| **PBE0** | 2.237 | 2.237 | 3.860 | 3.860 |
| **LC-ωPBEh** | 2.087 | 2.087 | 3.945 | 3.945 |
| **ωB97X-D** | 2.062 | 2.062 | 3.944 | 3.944 |
| **LC-ωPBE** | 2.087 | 2.087 | 3.945 | 3.945 |
| **TPSSh** | 2.245 | 2.245 | 3.791 | 3.791 |
| **M06-2X** | 2.174 | 2.174 | 3.940 | 3.940 |
| **M08-HX** | 2.224 | 2.224 | 3.936 | 3.936 |
| **ωM05-D** | 2.058 | 2.058 | 3.944 | 3.944 |
| **B97M-V** | 2.233 | 2.233 | 3.822 | 3.822 |
| **$r^2$SCANh** | 2.283 | 2.283 | 3.881 | 3.881 |
| **$r^2$SCAN0** | 2.271 | 2.271 | 3.970 | 3.970 |
| **ωB97X-D3** | 1.987 | 1.987 | 3.976 | 3.976 |
| **ωB97M-V** | 1.922 | 1.922 | 3.940 | 3.940 |
| **ωM06-D3** | 1.957 | 1.957 | 3.992 | 3.992 |
| **SCAN** | 2.143 | 2.143 | 3.828 | 3.828 |
| **MN15-L** | 2.304 | 2.304 | 3.949 | 3.949 |
| **M06-L** | 2.252 | 2.252 | 3.794 | 3.794 |
| **$r^2$SCAN** | 2.270 | 2.270 | 3.816 | 3.816 |
| **revM11** | 1.823 | 1.823 | 3.966 | 3.966 |
| **ωB97X-V** | 1.918 | 1.918 | 4.002 | 4.002 |
| **TPSS** | 2.171 | 2.171 | 3.653 | 3.653 |
| **M11** | 2.012 | 2.012 | 3.976 | 3.976 |
| **BP86** | 2.187 | 2.187 | 3.635 | 3.635 |
| **PBE** | 2.193 | 2.193 | 3.638 | 3.638 |
| **BLYP** | 2.164 | 2.164 | 3.609 | 3.609 |

**Table S5:** Vertical Excitation Energies (VEEs, in eV) associated with the Q- and B-bands of phot0-Zn calculated using DLPNO-STEOM-CCSD and TD-DFT with various exchange–correlation functionals with the geometry in S5.

| | $Q_y$ | $Q_x$ | $B_y$ | $B_x$ |
|---|---|---|---|---|
| **DLPNO-STEOM-CCSD** | 2.210 | 2.214 | 3.985 | 3.986 |
| **B2PLYP** | 2.267 | 2.267 | 3.746 | 3.746 |
| **ωB2PLYP** | 2.038 | 2.038 | 3.974 | 3.974 |
| **PW6B95** | 2.262 | 2.262 | 3.904 | 3.904 |
| **M06** | 2.186 | 2.186 | 3.844 | 3.844 |
| **B3LYP** | 2.256 | 2.256 | 3.855 | 3.855 |
| **MN15** | 2.188 | 2.188 | 3.932 | 3.932 |
| **M08-SO** | 2.154 | 2.154 | 3.965 | 3.965 |
| **CAM-B3LYP** | 2.141 | 2.141 | 4.009 | 4.009 |
| **PBE0** | 2.281 | 2.281 | 3.927 | 3.927 |
| **LC-ωPBEh** | 2.138 | 2.138 | 4.022 | 4.022 |
| **ωB97X-D** | 2.116 | 2.116 | 4.022 | 4.022 |
| **LC-ωPBE** | 2.047 | 2.047 | 4.023 | 4.023 |
| **TPSSh** | 2.290 | 2.290 | 3.851 | 3.851 |
| **M06-2X** | 2.227 | 2.227 | 4.030 | 4.030 |
| **M08-HX** | 2.276 | 2.276 | 4.022 | 4.022 |
| **ωM05-D** | 2.111 | 2.111 | 4.032 | 4.032 |
| **B97M-V** | 2.281 | 2.281 | 3.878 | 3.878 |
| **$r^2$SCANh** | 2.328 | 2.328 | 3.943 | 3.943 |
| **$r^2$SCAN0** | 2.316 | 2.316 | 4.040 | 4.040 |
| **ωB97X-D3** | 2.044 | 2.044 | 4.057 | 4.057 |
| **ωB97M-V** | 1.981 | 1.982 | 4.029 | 4.029 |
| **ωM06-D3** | 2.017 | 2.017 | 4.085 | 4.085 |
| **SCAN** | 2.195 | 2.195 | 3.888 | 3.888 |
| **MN15-L** | 2.358 | 2.358 | 4.025 | 4.025 |
| **M06-L** | 2.307 | 2.307 | 3.861 | 3.861 |
| **$r^2$SCAN** | 2.316 | 2.316 | 3.870 | 3.870 |
| **revM11** | 1.885 | 1.885 | 4.049 | 4.049 |
| **ωB97X-V** | 1.977 | 1.977 | 4.088 | 4.088 |
| **TPSS** | 2.223 | 2.223 | 3.709 | 3.710 |
| **M11** | 2.066 | 2.066 | 4.057 | 4.057 |
| **BP86** | 2.233 | 2.233 | 3.684 | 3.684 |
| **PBE** | 2.239 | 2.239 | 3.687 | 3.687 |
| **BLYP** | 2.211 | 2.211 | 3.656 | 3.656 |